\documentclass[12pt]{article}
\usepackage{amsmath}
\usepackage{amssymb}
\usepackage{epsfig}
\usepackage{cite}
\usepackage{color,colordvi}

\begin{document}
\vspace{0.01cm}
\begin{center}
{\Large\bf  Black Hole's  $1/N$ Hair} 

\end{center}

\vspace{0.1cm}

\begin{center}

{\bf Gia Dvali}$^{a,b,c,d}$\footnote{georgi.dvali@cern.ch} and {\bf Cesar Gomez}$^{a,e}$\footnote{cesar.gomez@uam.es}

\vspace{.6truecm}


{\em $^a$Arnold Sommerfeld Center for Theoretical Physics\\
Department f\"ur Physik, Ludwig-Maximilians-Universit\"at M\"unchen\\
Theresienstr.~37, 80333 M\"unchen, Germany}


{\em $^b$Max-Planck-Institut f\"ur Physik\\
F\"ohringer Ring 6, 80805 M\"unchen, Germany}

{\em $^c$CERN,
Theory Department\\
1211 Geneva 23, Switzerland}


{\em $^d$Center for Cosmology and Particle Physics\\
Department of Physics, New York University\\
4 Washington Place, New York, NY 10003, USA}

{\em $^e$
Instituto de F\'{\i}sica Te\'orica UAM-CSIC, C-XVI \\
Universidad Aut\'onoma de Madrid,
Cantoblanco, 28049 Madrid, Spain}\\

\end{center}


\begin{abstract}
\noindent  
 
{\small
 According to the  standard view classically black holes carry no hair, whereas quantum hair is at best exponentially weak.  
 We show that suppression of hair is an artifact of the semi-classical treatment and that in the quantum picture hair appears as an inverse mass-square effect.  
  Such hair is predicted in the microscopic  quantum description in which  a  black hole represents a self-sustained leaky Bose-condensate of $N$ soft gravitons.  
In this picture the Hawking radiation is the quantum depletion of the condensate. 
Within this picture we show that quantum black hole physics is fully compatible 
with continuous global symmetries and that global hair appears with the strength $B/N$, where 
$B$ is the global charge swallowed by the black hole.  For large charge this hair has 
dramatic effect on black hole dynamics.   Our findings can have interesting astrophysical consequences, such as   existence of black holes with large detectable baryonic and leptonic numbers.  }

\end{abstract}

\thispagestyle{empty}
\clearpage



%
%
%
 It is well known that in the semi-classical treatment  black holes  have no hair \cite{no-hair} and that they evaporate thermally \cite{hawking}.
The combination of these two properties leads to some "folk theorems", such as the
non-existence of exact global symmetries in the presence of  gravity.

   Intuitively, based on a common quantum field-theoretic sense, it is clear that the above 
   properties must be the result of the idealized  semi-classical treatment, and therefore they must be 
   corrected in the microscopic quantum-mechanical description. 
     However the explicit resolution of the issue  requires the existence of such quantum description.

We have recently painted  \cite{gia-cesar} such a quantum portrait of a black hole. 
 This picture is maximally simple in the sense that can be consistently quantified. 
 Namely, in our description, the black hole is a maximally overpacked Bose-condensate of 
very soft and very weakly-interacting gravitons.  What makes it different from 
other multi-particle systems (e.g., such as a  chunk  of ice), is precisely this overpacking. 
 In the absence of other charges, the occupation number of gravitons in the condensate, $N$,   is the only parameter.
 All the other characteristics are uniquely defined by $N$, as follows.   Occupation number  $=\,  N$, 
  constituent's wave-length $= \,  \sqrt{N} L_P$,   coupling stength $=  \,  1/ N$ and total mass   
  $\equiv \, M\, = \,  \sqrt{N}  {\hbar \over L_P} $. 
Here, the Planck length ($L_P$) and the corresponding Planck mass  ($M_P$) are defined  via Newton's coupling constant $G_N$ as, 
  $ L_P \, \equiv \, {\hbar \over M_P} \,  \equiv \, \sqrt{ \hbar \, G_N }\,$.
 These relations are the typical large-$N$ relations in  
't Hooft's sense\cite{tHooft} . 

The notion of graviton occupation number,
 $ N \, = \, M^2 G_N/\hbar \, = \, M^2/M_P^2  \, , $ 
as a quantum entity characterizing the level of classicality was introduced in \cite{class1, gia-cesar}.

    The above relations  are determined by the property that the black hole is a self-sustained 
    Bose-condensate. 
 The special  property of gravity is that such condensate exist for arbitrary $N$.     
 The reason is that the  graviton coupling depends on its wave-length and therefore can be self adjusted in order to balance the kinetic energy of an  individual graviton  versus  the 
 collective binding potential  of $N$ gravitons resulting into a self-sustained condensate for 
 arbitrary $N$.  
 
 A remarkable thing about  this condensate is that it is {\it leaky} whatever is the occupation number $N$!
  The leakiness of the condensate means that it continuously depletes due to quantum fluctuations that 
  excite some of the quanta above the escape energy.  This depletion is a fully-quantum 
  progenitor of what in  a semi-classical treatment is known as Hawking radiation. 
  However, as we have shown,  the exact thermality is only recovered in the semi-classical limit  which
 serves as a black hole analog of 't Hooft's planar limit.    
  This limit is an idealized approximation in which the important properties of real black holes become lost.  In particular,  the black hole mass becomes infinite,  whereas $G_N$ goes to zero.  However , for the finite values of the parameters, the radiation is never exactly thermal with the corrections being only $1/N$  (equivalently $1/M^2$)-suppressed. 
  
   The above is obvious from the quantum depletion law derived in \cite{gia-cesar}. This law reads, 
   \begin{equation}
   {\dot N} \, = -  {1 \over \sqrt{N} L_P}  \, + \, L_P^{-1} \,  {\mathcal O} (N^{-3/2}) \, , 
   \label{deplete}
   \end{equation}
 where dot stands for the time-derivative. 
 The leading term in this depletion comes from the two graviton scattering process, in which 
 one of them gains an above the escape energy  (see Fig. 1).   
  \begin{figure}[ht]
\begin{center}
\includegraphics[width=85mm,angle=0.]{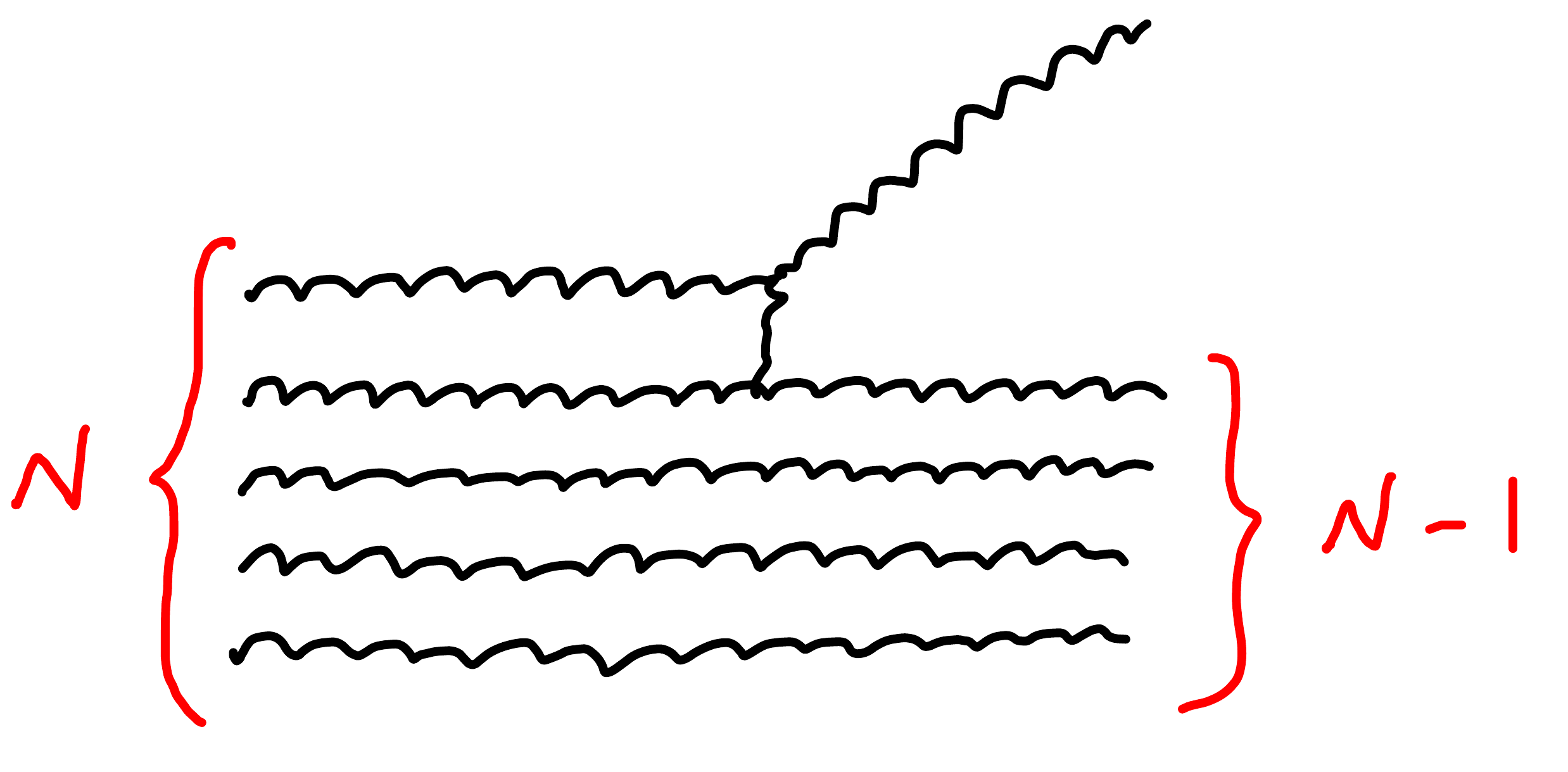}
\end{center}
\caption{Leading order process responsible for quantum depletion of graviton condensate.
}
\label{fig_safe}
\vspace{0.5cm}
\end{figure}
 Since the graviton-graviton coupling in the condensate 
 is $1/N$, the probability for any pair of gravitons to scatter  is suppressed by $1/N^2$, however this suppression  is compensated by a combinatoric factor  $\sim \, N^2$  counting the number of available graviton  pairs.  As a result,  the rate of the graviton emission from the condensate is simply given by 
 the characteristic energy of the process (inverse wave-length of gravitons) 
\begin{equation}
 \Gamma_{leakage} \, = \,   {1 \over  \sqrt{N} L_P}   \, + \,  L_P^{-1} \,  {\mathcal O} (N^{-3/2}) \, . 
\label{leakage}
\end{equation}
 The sub-leading corrections that are given by higher powers of $1/N$ in (\ref{leakage}) and 
 (\ref{deplete}) come both from the corrections  to the combinatoric factor in two graviton scattering as well as 
from  the diagrams involving more gravitons.   The important thing about these corrections is that  
they all scale with a common  single power of $L_P^{-1}$.  This is the key to why in the semiclassical limit  
the evaporation rate reproduces thermality.   More precisely the semi-classical limit is defined by the following double scaling limit, 
  \begin{equation}  
   N \, \rightarrow \, \infty, ~~~L_P\, \rightarrow \, 0\,,   ~~~ R\equiv \sqrt{N}L_P \, =\, {\rm finite}, ~~~\hbar \, = \, {\rm finite}\, ,
   \label{limit}
   \end{equation}
where we have kept fixed $R$ which has dimensionality of length.  
  This length  becomes in the semiclassical limit the Schwarzschild radius  of the black hole. 
 In this fashion we see that the notion of geometry only emerges in this limit.  Moreover only in this limit the 
   black hole depletion becomes exactly thermal and equivalent to Hawking's standard result. 
   To see this, let us rewrite 
 the depletion law (\ref{deplete}) in this limit, 
   \begin{equation}
   {\dot N} \, = -  {1 \over  R}  \, ,  
   \label{depleteexact}
   \end{equation}
 Or equivalently,  rewriting $N$ in terms of the black hole mass we get 
    \begin{equation}
   {\dot M} \, = -  {\hbar  \over  R^2}  \, . 
   \label{depletemass}
   \end{equation}
 The latter expression is nothing but the Stefan-Boltzmann law for a black hole with  Hawking temperature given by  $T \, =  \, {\hbar  \over R}$.  It is very important to 
 stress that what allows us to convert the depletion law (\ref{deplete}) into a Stefan-Boltzmann law
 is the self-similarity of the depletion with respect to $N$.  This is a very special property of the 
 self-sustained graviton condensate, which is not exhibited by other  unstable systems  (e.g., such as  alpha-decay). 
   
   The above clearly shows that thermality is a property only emerging in a planar (semi-classical) limit 
   which is never obeyed by any realistic black holes, since $M$, $G_N$ and $\hbar$ are  
  finite in nature.  For any real black hole the thermality is violated by corrections that scale 
 with inverse power of the occupation number or equivalently of the black hole mass.   
 Besides, violation of the black hole thermality is closely linked with evading the  black hole no-hair theorem.   
   
 
   It is well established\cite{no-hair}  that classical black holes can only carry hair under the charges that result into the 
   Gaussian gauge fluxes at infinity \footnote{Here we are not interested in a so-called secondary hair, which 
   can result by a trivial sourcing of some massive field by the primary massless gauge hair.}. 
  Such are the charges that source the massless gauge fields, e.g., an  
   electric charge.  
  It was noticed \cite{qh1} that in very special circumstances quantum-mechanically the black holes  
 can carry topological (Aharonov-Bohm  type) hair even in the absence of the massless gauge fields.  This happens whenever a discrete subgroup of the gauge group remains 
 un-Higgs.
 Such hair can be detected by performing global topologically-non-trivial Aharonov-Bohm measurements.  
  However,  it has been shown \cite{qh2} that  even in such circumstances the locally 
  detectable hair (e.g., such as an electric field) is  exponentially suppressed and is given by, 
  \begin{equation}
   {\rm Electric \, ~hair} \, \propto  \,   {\rm e}^{- {S_{string} \over \hbar} } \, {\rm e}^{- {\mu R \over \hbar} } \, ,   
  \label{exp}
  \end{equation}
 where $\mu$ is the mass of the gauge field.  The last exponent in  this expression reminds 
  the usual Yukawa screening of the electric charge at distances larger than the Compton wave-length of the gauge field.  Such screening would take place already at the classical level for a 
 static charge in a  gauge theory that is in the Higgs phase.   The non-perturbative semi-classical 
 nature of the hair is revealed by the first exponent.  This exponent  indicates that hair comes from the  virtual processes in which  a cosmic string lassoes the black hole.  $S_{string}$ stands for the 
 Euclidean action describing such a transition.  
  Even such a negligible hair is only  associated with very special (discrete gauged)  charges.  
    In the standard semi-classical treatment,  all the other information swallowed by a black hole carries no hair.   
    
    We shall now demonstrate that the absence of hair is an artifact of the  above-discussed 
  semi-classical  (planar) limit. In reality there is a much stronger source of the hair which is only 
  power-law suppressed by the ratio of the charge to the black hole mass$^2$.    
   Moreover such hair is present, even if the charge in question is global.

  Intuitively, the necessary presence of black hole hair in the  quantum picture is already expected from the violation of  thermality by $1/N$ corrections in (\ref{deplete}). However,  let us make the argument more concrete 
  and precise. 
  
  In order to tag the information,  consider an elementary particle  $\phi$ that carries a conserved charge 
  not associated with any long-range massless gauge field.  Fo example, a $Z_2$-symmetry could do. 
  In the standard semi-classical treatment,  since black hole carries no hair,   such a charge
  cannot be respected by a black hole, unless $Z_2$ is promoted into a gauge symmetry.  
   The absence of hair then is used as an argument that non-gauge global symmetries cannot be respected by black holes. 
 Since, we are going to show that the hair inevitably appears at the order $1/N$, the pre-condition 
 about gauging the symmetry becomes unimportant.   As said above, even if symmetry is gauged the 
 hair appearing in the standard treatment is exponentially suppressed (see (\ref{exp})). This will 
 be a  totally unimportant correction to our result,  since the quantum effect we are going to show 
 becomes apparent at much stronger level. 
 
  Let us perform the following thought experiment.  Imagine that a single $\phi$ quantum is a part of the black hole condensate.   Then such a condensate carries a single unit of the $Z_2$-charge.  
 Let us ask to which order in $1/N$ such a particle will reveal itself in long-wavelength 
 measurements.    In order to see this consider an observer (Alice) that is trying to 
 probe the $Z_2$-charge of the condensate using some  long-wavelength external probe. 
  What are her chances to discover that the condensate contains the $\phi$-quantum? 

 Quantum-mechanically  $\phi$ can reveal its existence to an outside probe through 
 a virtual process in which $\phi$ (virtually) depletes by scattering-off one of the $N-1$ gravitons, then scatters 
 off an external probe and returns back to the condensate.  
 It is obvious that the rate of this process is $1/N$-suppressed as compared to the rate of the depletion of
 radiation (\ref{deplete}).  Indeed, the above-described process of the appearance of the $\phi$-hair is very similar to depletion.  All we have to do is to replace the escaping graviton line in the depletion Feynman diagram by a $\phi$ line and add a vertex describing an  interaction with the external probe prepared by Alice (see Fig.2).  
   \begin{figure}[Ht]
\begin{center}
\includegraphics[width=85mm,angle=0.]{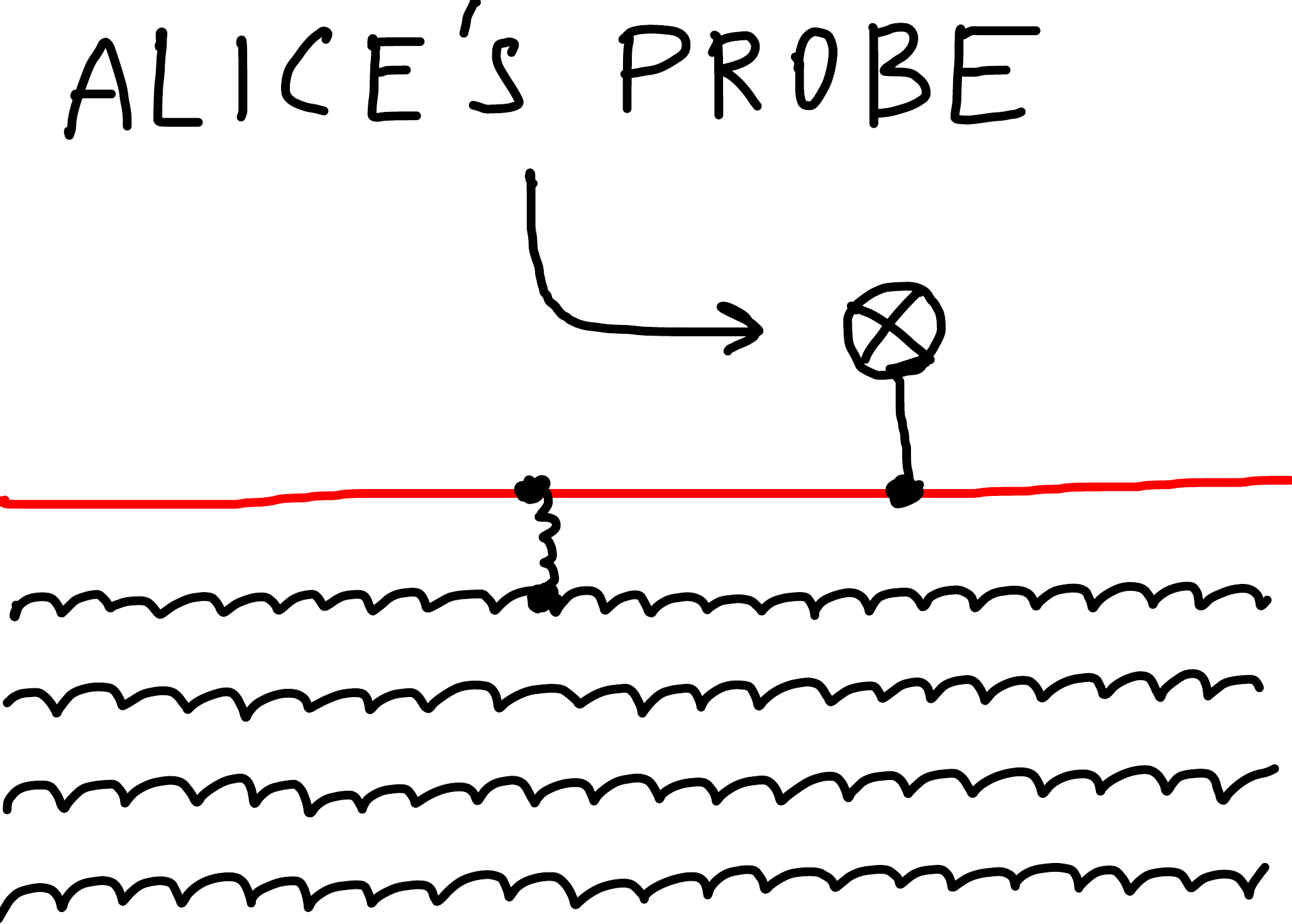}
\end{center}
\caption{Leading order process responsible for the detection of quantum hair by Alice. 
 The red solid line corresponds to $\phi$-particle.  The interaction with the Alice's probe 
 is encoded in the crossed vertex.}
\label{fig2}
\vspace{0.5cm}
\end{figure}
   The latter vertex  contributes to the probability with some fixed number 
 (call it $\alpha_{probe}$) which is  independent of $N$ and is determined by the properties of Alice's detector. 
 
 The probability coming from the graviton scattering vertex  is exactly the same as in the thermal depletion case, and gives factor $1/N^2$. But, the combinatorics factor is now $N$ instead of  $N^2$, since one of the 
 participants must be $\phi$.   Thus, the rate at  which Alice measures the $\phi$-hair  is only suppressed as, 
 \begin{equation}
 \Gamma_{hair} \,  =  \,   \alpha_{probe} {1 \over  N^{3/2} L_P}   \, + \,  L_P^{-1} \,  {\mathcal O} (N^{-{5/2}}) \, . 
\label{ratehair}
\end{equation}
   Therefore, in  the leading order in $1/N$ we have the relation, 
 \begin{equation}
 \Gamma_{hair} \,  = \Gamma_{leakage} \, {\alpha_{probe}  \over  N} \, ,   
\label{thermalhair}
\end{equation}
 which shows that the hair is $1/N$ effect relative to the thermal depletion.  
This is why in  the semi-classical limit (\ref{limit}) the hair disappears, and if $Z_2$ is not gauged Alice measures nothing. If $Z_2$ is gauged, the contribution (\ref{exp}) remains, but of course is negligible. 
 Thus the planar limit is in full agreement with the known no-hair theorems, but also shows that these theorems are not applicable to real black holes with finite $N$. 
 
   So far we have shown that the global  hair appears as $1/N$-effect per single quantum carrying the global charge.  We now show that  this effect can have a dramatic influence on arbitrarily large 
   black holes when the number of globally-charged quanta swallowed by a black hole  is large. 
   This finding has immediate implication for evading the standard argument of violating 
   global charges by the black hole physics.

 In order to see this,  let us generalize  our reasoning  to the case when a black holes carries 
more than one unit of some conserved global charge.  For this we need to enhance 
$Z_2$-symmetry. For example we can promote it into a fully-continuous global symmetry 
$U(1)_B$, with a conserved charge $B$, to which we should generically refer as "baryon number".  We can repeat the same analysis for a black hole
that in the condensate stores $B$-units of the global charge in form of $\phi$-quanta each carrying 
one unit.   It is obvious that the rate (probability per unit time)  for Alice to detect such a
charge is now enhanced by factor $B$.  For the case  $N \, \gg \, B$ we will have,  
 \begin{equation}
 \Gamma_{hair} \,  = \,   \alpha_{probe} { B \over  N^{3/2} L_P}  \left (1 \, + \, \,  {\mathcal O} (B/N) \right )  \, . 
\label{rateB}
\end{equation}
From here we can write a depletion law for the global charge, 
   \begin{equation}
   {\dot B} \, = -  {1 \over \sqrt{N} L_P} {B \over N}  \left (1 \, + \,  \,  {\mathcal O} (B/N)  \right )\, .  
   \label{depleteB}
   \end{equation}
Correspondingly,  in the presence of the baryonic charge,   there will be corrections to the graviton depletion law of  $N$ of the order $B/N$.    It is illuminating to rewrite equation (\ref{depleteB}) in the  form,  
  \begin{equation}
  { {\dot B} \over {\dot N}} \, =  \, {B \over N} \,  \left (1 \, + \,  \,  {\mathcal O} (B/N)  \right )\, .  
   \label{depleteBN}
   \end{equation}
  Or in terms of the black hole mass, 
  \begin{equation}
  { {\dot B} \over {\dot M}} \, =  \, {B \over M} \,  \left (1 \, + \,  \,  {\mathcal O} (B\hbar^2 /M^2L_P^2)  \right )\, .  
   \label{depleteBM}
   \end{equation}

 This form tells us that  the effect of baryonic hair relative to the leading order depletion is a 
 $B/N$-effect.  This fact makes it immediately clear  why all our findings are hidden in the  
semi-classical limit.  Indeed,  taking the limit  (\ref{limit}) with $B$ fixed, we get that 
the depletion becomes thermal and the effect of the hair vanishes.

 Our finding re-opens the question of compatibility of the conserved global quantum numbers 
with the black hole physics. Having a fully quantum mechanical description of the black hole hair, we can now trace were the standard argument fails. 

The standard believe that  the black holes should not respect global symmetries, such as
baryon and  lepton numbers, is based  on  the assumption of exact thermality  and non-existence of hair. 
  In order to see more precisely where is the loophole,  let us make the argument more rigorous.
  Let us perform the following thought experiment \footnote{ We could trace a version of this  thought experiment  to as early as \cite{witten}.   A similar  argument was made quantitative and was used to put a bound on dimensionality of maximal representation of the  global symmetry group 
  in \cite{species, bs}.}.
Let us assume, that the  semi-classical statements about thermality and absence of hair are valid for the black holes larger than a certain critical mass $M_c$.  
Then, we can always  prepare an arbitrarily heavy  black hole and endow it with an arbitrarily large 
global charge $B$, e.g., the baryon number. 
 Imagine now that Alice is monitoring evaporation of such a black hole.  Alice knows that an amount 
 $B$ of an initial baryonic charge went onto a black hole, but she has no way to measure it 
 during the evaporation process, due to exact thermality and the absence of hair. 
 If $B$ is conserved, then either  Alice should sooner or later retrieve it or the charge should remain 
 permanently stored within the final remnants.  Strictly speaking these two options are equivalent 
 since if $B$ is conserved Alice should be able to consistently attribute it to some final states, 
 never mind how she calls them,  particles or the black hole remnants.
   
    The crucial assumption of the semi-classical treatment is, that by default any retrieval can only take place after the black holes shrink beyond  the fix mass $M_c$, since before that nothing can happen due to the absence of hair and thermality. In other words, the semi-classical argument assumes that 
 because a black hole  with $M \, > \, M_c$ carries no hair the critical mass  $M_c$ is independent of  initial $B$.       
   This inevitably creates a problem, since $B$ can be arbitrarily large. So the only possibility 
   to reconcile arbitrariness of initial $B$ with  fixed value of $M_c$ is to sacrifice conservation of 
   $B$.   Hence the standard conclusion is that black holes must violate global symmetries or equivalently that global symmetries don't exist in the presence of gravity.  
 The place were this semi-classical reasoning goes wrong is that it reconciles 
 the absence of hair and exact thermality  with the possibility of  reaching ,by thermal evaporation,  a {\it fixed critical mass} $M_c$. 

 However, according to what we have shown in this note,  the above two notions are incompatible. 
  Either $M$ (and  $N$) is infinite and then black hole is eternal, or $M$ (and $N$) is finite and 
then  hair exists as $B/N$-effect and is fully manifest.   In both cases, there is no conflict with conservation of the baryonic charge. 
  The truth about the evaporation process is encoded in equation (\ref{depleteBN}). 
  According to this equation for Alice there is no mystery,  since she can continuously monitor the baryonic charge of a black hole.  For her the baryonic charge carried by a black hole is as well-defined as the  baryon number of  a  neutron.   
 
 
   In more precise terms, according to (\ref{depleteBN}), a  large black hole initially endowed with a relatively-small baryonic charge,  $B\, \ll \, N$,  
 will  deplete at the beginning as approximately-thermal, emitting very little baryonic 
charge per graviton.   Baryon-to-graviton emission rate will be of order $B/N$.  

  In this regime, the equations (\ref{deplete}) and (\ref{depleteBN}) can be integrated explicitly
and we get 
\begin{equation}
N(\tau) \, = \, (\tau_* \, - \, \tau)^{{2\over 3}} \, ~~~\,  B(t)\, = \, \left (1 - {\tau \over \tau_*} \right )^{{2\over 3}}
\, B(0) \, , 
 \label{solution}
 \end{equation}
where  the parameter  $\tau \, \equiv \, {2\over 3} {t \over L_P}$ measures in Planck units the time since the start of depletion, and the initial value of the occupation number is given by 
$N(0) \, =\, \tau_*^{{2\over 3}}$.   The above solutions are valid as long as 
$\tau\, \ll \, \tau_*$. They explicitly indicate that $B$ can start changing substantially  only after 
$\tau$ becomes of order $\tau_*$, that is, after the black hole has depleted of order half of its 
graviton budget.  
   So the baryonic charge exhibits a clear tendency to catch up with $N$. 
 This tendency continues until $B\, \sim \, N$.   
  At this point deviation from thermality   will become 
 critical and  higher order corrections in $B/N$ have to be taken into the account. 
  This point defines the critical mass $M_c(B)$, as 
 \begin{equation}
 M_c(B) \, \sim  \, \sqrt{B} \, \hbar /L_P\, . 
 \label{mcritical}
 \end{equation}  
For such a black hole  the  effect of the baryonic hair  becomes hundred percent important, and 
the baryonic charge of the black hole is fully transparent to Alice.  

 The equation (\ref{mcritical})   is very instructive for confronting the semiclassical reasoning with the full quantum one, since the same equation appears in the semi-classical treatment \cite{species, bs}, but with a completely different  meaning.  
    In semi-classical argument one treats the scale $M_c$ as a fixed fundamental scale 
 above which thermality and no-hair properties must be obeyed.   As a result the equation (\ref{mcritical}) acquires the meaning of the  bound that constraints, in terms of this fundamental scale
 $M_c$,  the maximal allowed value of global charge $B$ (equivalently the maximal allowed dimensionality of the representations of the global symmetry group)  in the theory  \cite{species,bs}.   This automatically excludes the continuous global symmetries, which would allow $B$ to be arbitrarily large. 
  Our quantum picture shows that  instead 
 $M_c(B)$ is not a fixed scale, but a characteristics of a particular black hole with global charge $B$.   So the value of $B$ is not limited by any consistency considerations, since $M_c(B)$ can take arbitrarily large values  in a given theory.  As a result the continuous global symmetries are  permitted to coexist with black hole physics.

This situation is reminiscent of Page's findings on black hole retrieval of information \cite{Page}. Page's main point lied on using a generic quantum mechanical model of a black hole as a system with a finite dimensional Hilbert space. In our portrait the dimension will be of $O(2^N)$ and it will diminishes during the evaporation process at the rate determined by the depletion equation. The information encoded in the $n$ radiated quanta is for $n<<\frac{N}{2}$ exponentially small and the "depletion" of information in this initial state of the evaporation will be
\begin{equation}
\frac{dI}{dt} \, \propto \,  \frac{1}{\sqrt{N}L_P} e^{-N}
\end{equation}
Note that in the semiclassical $N=\infty$ limit no information is returned in the radiated quanta. The situation changes when $n>>\frac{N}{2}$. In this regime the black hole starts to retrieve information almost at the depletion rate. Similarly in the case of a black hole endowed with some baryon number $B$ when $B<<N$ the quantum depletion of baryon number is very much suppressed disappearing in the strict semi-classical limit, however when $B \sim N$ i.e when $B$ is of the order of the constituent gravitons, quantum depletion of baryon number is in principle of the same order as quantum depletion. 

 Before ending let us point out that a very interesting question is of course to unveil what happens to black holes charged with some global charge beyond the point $M_c(B)$. This issue will not be discussed 
  here and it very much depends on how the self-sustainability of the condensate works when the number of constituent gravitons and baryons is of the same order. Here we will just briefly comment that it is plausible that in this case the black hole depletion (nearly) stops 
 and it gets stabilized in some sort of extremal state, but without any long-range gauge field.   
  Such globally-extremal black holes, even if  not exactly stable but nevertheless long-lived,  can have very interesting cosmological and astrophysical 
   consequences.  
   
  In the above example we have considered an oversimplified model of baryons 
with massless quanta interacting only gravitationally and we have ignored the effects of other interactions.      
   In realistic applications,  if,  for example, $B$ is an ordinary baryon number,
 one has to take into the account the effect of the nuclear and electroweak  forces.    
 In the absence of such forces    
  a critical mass  
 baryonic black hole would  have a size of a neutron, but will carry $B \, = \, 10^{38}$ units of baryonic charge
 and weight approximately $10^{15}$g.   In reality the corrections from other forces will be very important and can reveal the effects of the baryonic hair for much bigger black holes. 
   This question  deserves  a further investigation because of its obvious observational importance.

  Of course, there may exist independent sources for  breaking  global symmetries, such as, for example, anomalies.  About this we have nothing new to say.   The purpose of the present paper 
  was to eliminate the main suspects that are usually invoked against the existence 
 of global symmetries.    In our quantum picture black holes can  coexist with continuous global symmetries,  while accommodating all the known properties of the semi-classical limit.  
 The result of this coexistence is the possibility of having black holes with fully fledged 
 baryonic and leptonic charges, that  will have macroscopic effect.

%
%


\section*{Acknowledgements}

It is a pleasure to thank Costas Bachas, Slava Mukhanov and Alex Pritzel  for discussions, as well as 
John Iliopoulos and  Costas Kounnas  for hospitality at  Ecole Normale Superieure  where 
part of this work was presented. 
The work of G.D. was supported in part by Humboldt Foundation under Alexander von Humboldt Professorship,  by European Commission  under 
the ERC advanced grant 226371,   by TRR 33 \textquotedblleft The Dark
Universe\textquotedblright\   and  by the NSF grant PHY-0758032. 
The work of C.G. was supported in part by Humboldt Foundation and by Grants: FPA 2009-07908, CPAN (CSD2007-00042) and HEPHACOS P-ESP00346.

\end{document}